\documentclass[journal]{IEEEtran}

\ifCLASSINFOpdf
\usepackage[pdftex]{graphicx}
\graphicspath{{../pdf/}{../jpeg/}}
\DeclareGraphicsExtensions{.pdf,.jpeg,.png}
\else
\usepackage[dvips]{graphicx}
\graphicspath{{../eps/}}
\DeclareGraphicsExtensions{.eps}
\fi
\usepackage{cite}
\usepackage{algorithmic}
\usepackage{amssymb}
\usepackage{textcomp}

\usepackage{amsmath}
\usepackage{amsfonts}
\usepackage{flushend}
\usepackage{amssymb}
\usepackage{makecell}
\usepackage{booktabs}
\ifCLASSOPTIONcompsoc

\usepackage[caption=false,font=normalsize,labelfont=sf,textfont=sf]{subfig}
\else
\usepackage[caption=false,font=footnotesize]{subfig}
\fi
\usepackage{graphicx}

\usepackage{epsfig}
\usepackage{epstopdf}
\usepackage{color}
\usepackage{bm}

\let\emph\textit
\usepackage{fixltx2e}

\usepackage{multicol}  
\usepackage{multirow}

\def\BibTeX{{\rm B\kern-.05em{\sc i\kern-.025em b}\kern-.08em
		T\kern-.1667em\lower.7ex\hbox{E}\kern-.125emX}}

\begin{document}
	
\title{ Theoretical Analysis of Near-Field MIMO Channel Capacity and Mid-Band Experimental Validation }

%
%
%

\author{Haiyang~Miao, \IEEEmembership{Graduate Student Member,~IEEE}, Jianhua~Zhang, \IEEEmembership{Senior Member,~IEEE}, Pan~Tang, \IEEEmembership{Member,~IEEE}, Heng Wang, \IEEEmembership{Member,~IEEE}, Lei Tian, \IEEEmembership{Member,~IEEE}, Guangyi Liu, \IEEEmembership{Member,~IEEE}
	

\thanks{This work was supported by National Key R\&D Program of China (2023YFB2904802), National Natural Science Foundation of China (62201086, 62101069), BUPT-CMCC Joint Innovation Center. (\itshape{Corresponding author: Jianhua Zhang.})}
\thanks{H. Miao, J. Zhang, P. Tang, H. Wang and L. Tian are with State Key Laboratory of Networking and Switching Technology, Beijing University of Posts and Telecommunications, Beijing 100876, China (e-mail: hymiao@bupt.edu.cn; jhzhang@bupt.edu.cn; tangpan27@bupt.edu.cn; hengwang@bupt.edu.cn; tianlbupt@bupt.edu.cn).}
\thanks{G. Liu is with Future Research Laboratory, China Mobile Research Institute, Beijing 100053, China (e-mail: liuguangyi@chinamobile.com).}
}


\markboth{Journal of \LaTeX\ Class Files,~Vol.~14, No.~8, August~2021}%
{Shell \MakeLowercase{\textit{et al.}}: A Sample Article Using IEEEtran.cls for IEEE Journals}

%

\maketitle

\begin{abstract}	
	
	 With the increase of multiple-input-multiple-output (MIMO) array size and carrier frequency, near-field MIMO communications will become crucial in 6G wireless networks. Due to the increase of MIMO near-field range, the research of near-field MIMO capacity has aroused wide interest. In this paper, we focus on the theoretical analysis and empirical study of near-field MIMO capacity. First, the near-field channel model is characterized from the electromagnetic information perspective. Second, with the uniform planar array (UPA), the channel capacity based on effective degree of freedom (EDoF) is analyzed theoretically, and the closed-form analytical expressions are derived in detail. Finally, based on the numerical verification of near-field channel measurement experiment at 13 GHz band, we reveal that the channel capacity of UPA-type MIMO systems decreases continuously with the communication distance increasing. It can be observed that the near-field channel capacity gain is relatively obvious when large-scale MIMO is adopted at both receiving and transmitter ends, but the near-field channel capacity gain may be limited in the actual communication system with the small antenna array at receiving end. This work will give some reference to the near-field communication systems.

\end{abstract}


\IEEEpeerreviewmaketitle	
	
\section{Introduction}

\par In the 5G era, in order to meet the high capacity demand of communication, 3D MIMO, which makes full use of elevation domain to achieve flexible strategies, has been studied [1]. The 3D channel capacity bound was derived, which showed a clear relationship between 2D and 3D channel capacity [2], and the influencing factors of 3D channel capacity increase were analyzed. At present, the research on sixth-generation (6G) has been carried out on a global scale [3-4]. To accomplish a multi-fold increase in communication capacity for 6G, extremely large-scale multiple-input-multiple-output (XL-MIMO) is considered one of the potential key technologies [5-8]. Compared with traditional MIMO technology, the near-field communication becomes a key scenario due to the large physical size of XL-MIMO. In [9], the comparisons between systems under different practical configurations were well carried out through simulations. The simulation results comprehensively revealed that the capacity of discrete linear array converged to that of continuous aperture linear array at mmWave band [10]. The antenna array was designed in combination with the frequency modulation mode to achieve accurate channel capacity estimation [11]. The results demonstrated that the low-complexity subarray-wise method has relatively good performance [12]. In [13], the Shannon point-to-point capacity of MIMO reflecting intelligent surfaces systems was evaluated.

\par However, there are some views in the industry that insufficient near-field gain is seen in the real communication systems deployed with base station array, whose possible reasons need to be explored. Among, a key question is whether there is a significant gain in near-field MIMO channel capacity. The effective degree of freedom (EDoF) of a MIMO system represents its equivalent number of independent single input single output (SISO) systems [14], which makes it easier to characterize the performance of MIMO systems. The uniform planar array (UPA) is widely used in commercial communication systems. Therefore, this letter analyzes the factors affecting the variation of UPA-type MIMO near-field capacity performance based on EDoF. It is worth noting that most of the near-field channel capacity analysis is based on the simulation verification, so more discussion and analysis are needed through measurement experimental results.

\par On the other hand, the near-field range depends not only on the physical size of the antenna array, but also on the operating frequency. Currently, the new mid-band (6-24 GHz) has attracted extensive attention from academia and industry for 6G [15-17]. In December 2023, the 3rd Generation Partnership Project (3GPP) Rel-19 has determined the need to study the near-field channel in 7-24 GHz band [18]. Therefore, the mid-band is used as the channel measurement frequency to discuss the gaps and problems mentioned earlier in this letter. Firstly, the near-field channel model is characterized from the electromagnetic perspective. Then, with the UPA-type MIMO, the EDoF and channel capacity are analyzed theoretically and studied empirically. The closed-form analytical expressions of channel capacity are derived in detail. Thirdly, the empirical study of near-field channel measurement is carried out in the 13 GHz band. Meanwhile, we discuss and verify the near-field MIMO channel capacity of different aperture antenna arrays.


\section{Theoretical Analysis of Near-field Channel Model and Performance}

\begin{figure*}[!htbp]
	\setlength{\abovecaptionskip}{0.1 cm}
	\centering
	\includegraphics[width=0.7\textwidth]{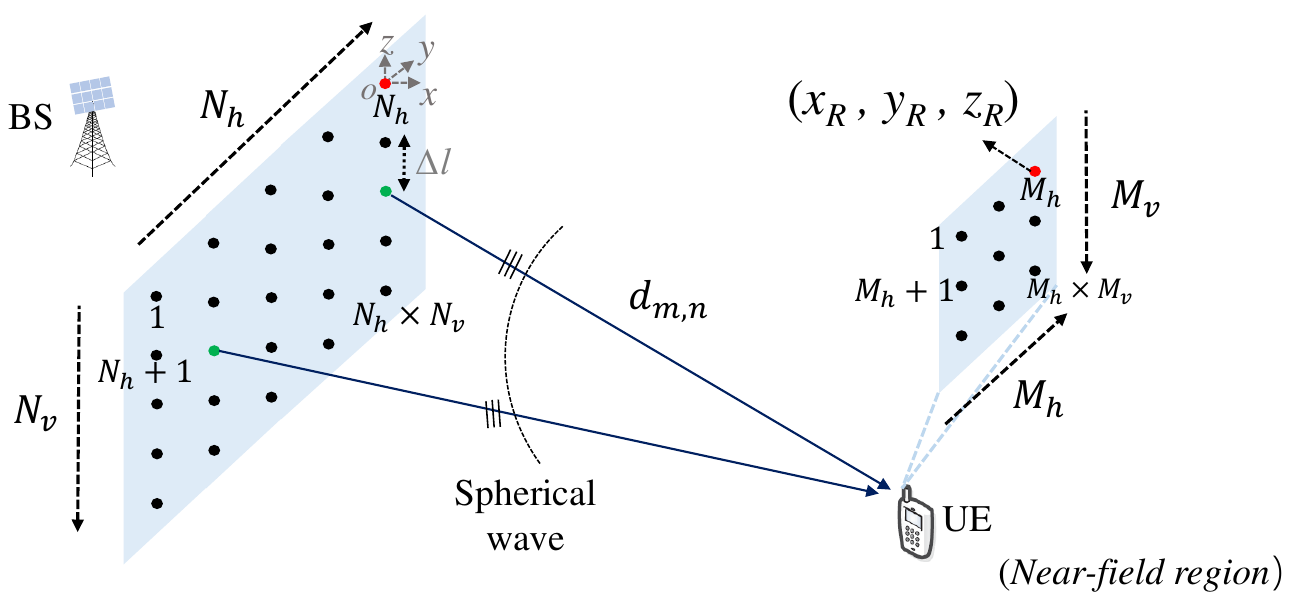}
	\caption{Near-field MIMO channel propagation model.}
	\label{fig:Figure1}
\end{figure*}

\par With the significantly increased aperture of MIMO, near-field communication will become critical. Due to the channel sparsity, the medium and high frequency communication is mostly line-of-sight (LOS) environment [15, 19]. The near-field and LOS propagation lead to new characteristics of MIMO. Due to the large physical size of the antenna array, the different elements on the array will see large and small scale channel fading characteristics fluctuate. The fluctuation of these channel characteristics will not only lead to the change of near-field channel model, but also affect the performance of near-field wireless communication system. Therefore, it is necessary to characterize and analyze the channel model and system performance in the near field.

\par As shown in Fig. 1, considering that the phase difference is the largest when the electromagnetic wave is vertically incident, the normal direction of the planar array antennas can be set parallel to the X-axis at the transmitting and receiving ends. For the transmitting array at the base station, the antenna array is deployed in the area of $N=N_h \times N_v$. $n \in [1, N]$ is the antenna array element index. Therefore, the coordinates of the $n$-th element on the transmitting array is expressed as

\begin{equation}
\begin{split}
T_n & = [x_n, y_n, z_n]\\
&=[0, (i(n)-N_h)\Delta l, (v(n)-N_v)\Delta l],
\end{split} 
\end{equation}
where $i(n)=\textup{mod}(n-1, N_h)$, and $v(n)=\textup{floor}((n-1)/N_h)$ represent the index of the $n$-th transmitting antenna in the horizontal dimension and the vertical dimension, respectively. The $\textup{mod}(.,.)$ is the mode operation, and the $\textup{floor}(.)$ is the round down operation. The element spacing of the UPA is $\Delta l= \lambda_0/2$. $\lambda_0$ is the wavelength.

\par Then, considering the position of the user is random, assume that the center position of the receiving array is $(x_R, y_R,z_R)$. For the receiving array at the user end, the antenna array is deployed in the area of $M=M_h \times M_v$. $m \in [1, M]$ is the antenna array element index. The coordinates of the element $m$ on the receiving array can be expressed as

\begin{equation}
\begin{split}
R_m & = [x_m, y_m, z_m]\\
& = [x_R, y_R+(k(m)-M_h)\Delta l, \\
& z_R+(w(m)-M_v)\Delta l],
\end{split} 
\end{equation}
where $k(m)=\textup{mod}(m-1, M_h)$, and $w(m)=\textup{floor}((m-1)/M_h)$ represent the index of the $m$-th receiving antenna in the horizontal dimension and the vertical dimension, respectively.

\subsection{Channel Model}

\par Due to the $N$ array elements at the transmitting positions $T_n$, the electric field intensity $E$ at the receiving positions $R_m$ can be written as

\begin{equation}
\begin{split}
E_m & = \dfrac{1}{4\pi}\sum_{n=1}^{N} \dfrac{exp(-jk_0d_{m,n})}{d_{m,n}}a_n= \sum_{n=1}^{N}h_{m,n}a_n ,
\end{split} 
\end{equation}
where $k_0= 2\pi/\lambda_0$ is the wave number. $a_n$ is the signal complex amplitudes. $h_{m,n}$ is the Green's function with the Tx or Rx positions, which is the channel gain between the $n$-th antenna (subarray) of the Tx and the $m$-th antenna (subarray) of the Rx. Then, the channel matrix $\bm{H}\in \mathbb{C}^{M\times N}$ can be defined as

\begin{equation}
\begin{aligned}
\bm{H}
=
\begin{bmatrix}
h_{1,1} & h_{1,2} & \cdots & h_{1,N}\\
h_{2,1} & h_{2,2} & \cdots & h_{2,N}\\
\vdots &  \vdots  &  \ddots &  \vdots \\
h_{M,1} & h_{M,2} & \cdots & h_{M,N}\\
\end{bmatrix}_{M\times N},
\end{aligned} 
\end{equation}
where $h_{m,n}$  can be expressed as

\begin{equation}
\begin{split}
h_{m,n} & = \dfrac{exp(-jk_0d_{m,n})}{4\pi d_{m,n}}.
\end{split} 
\end{equation}

\par Among, $d_{m,n}$ can be expressed as

\begin{equation}
\begin{split}
d_{m,n} & = |R_m- T_n| \\
&=  \sqrt{(x_m-x_n)^2 + (y_m-y_n)^2 +(z_m-z_n)^2} \\
&=   x_R \sqrt{1 + (\dfrac{y_m-y_n}{x_R})^2 +(\dfrac{z_m-z_n}{x_R})^2},
\end{split} 
\end{equation}
where the $d_{m,n}$ is the distance between the Tx array element and the Rx array element. According to Taylor's expansion $\sqrt{1+x}\cong1+\dfrac{x}{2}$, the $d_{m,n}$ can be simplified as

\begin{equation}
\begin{split}
d_{m,n} & = x_R \sqrt{1 + (\dfrac{y_m-y_n}{x_R})^2 +(\dfrac{z_m-z_n}{x_R})^2} \\
&\approx x_R + \dfrac{(y_m-y_n)^2}{2x_R} + \dfrac{(z_m-z_n)^2}{2x_R}  \\
&= x_R + \dfrac{(y_m-y_n)^2 + (z_m-z_n)^2}{2x_R} .
\end{split}
\end{equation}

\subsection{EDoF}

\subsubsection{Expression}

\par In the near-field region, the singular values are different due to the influence of spherical wave, the concept of EDoF can be used to characterize the channel capacity. The concept and expression of EDoF were introduced in [14, 20]. The increased degree of freedom can be exploited as an additional spatial multiplexing gain, which provides the new possibility for capacity enhancement. The analytical expression of EDoF is first discussed here, which is expressed as

\begin{equation}
\begin{split}
\beta & = (\dfrac{tr(\bm{R})}{||\bm{R}||_F})^2 = \dfrac{(tr(\bm{R}))^2}{tr(\bm{R}^2)},
\end{split} 
\end{equation}
where $ ||\cdot ||_F $ denotes the Frobenius norm operation, and $tr(\cdot)$ represents the trace operator. $\bm{R} \in \mathbb{C}^{N\times N}$ is the channel correlation matrix, which is given by

\begin{equation}
\begin{split}
\bm{R} & = \bm{H}^\mathrm{H}  \bm{H},
\end{split} 
\end{equation}
where $ \left(\cdot \right)^\mathrm{H} $ denotes the conjugate transpose operation.

\subsubsection{Closed-Form Model}

\par In order to obtain the closed-form expression of EDoF, the expression of the channel correlation matrix is first given:

\begin{equation}
\begin{split}
\bm{R} & = \bm{H}^\mathrm{H}  \bm{H} \\
& =\begin{bmatrix}
\sum_{m=1}^{M}h^*_{m,1}h_{m,1} &  \cdots & \sum_{m=1}^{M}h^*_{m,1}h_{m,N}\\
\sum_{m=1}^{M}h^*_{m,2}h_{m,1} &\cdots & \sum_{m=1}^{M}h^*_{m,2}h_{m,N}\\
\vdots &  \ddots &  \vdots \\
\sum_{m=1}^{M}h^*_{m,N}h_{m,1} & \cdots & \sum_{m=1}^{M}h^*_{m,N}h_{m,N}\\
\end{bmatrix}_{N\times N}
\end{split} 
\end{equation}

\par Then, $\bm{R}(n_1, n_2)$ can be expressed as

\begin{equation}
\begin{split}
\bm{R}(n_1, n_2) & = \sum_{m=1}^{M}h^*_{m,n_1}h_{m,n_2}\\
& = \sum_{m=1}^{M}\dfrac{exp(jk_0(d_{m,n_1}-d_{m,n_2}))}{(4\pi)^2 d_{m,n_1}d_{m,n_2}}.
\end{split} 
\end{equation}

\par Thus, the trace of the matrix $\bm{R}$ can be derived as

\begin{equation}
\begin{split}
tr(\bm{R})  = \sum_{n=1}^{N}\bm{R}(n,n),
\end{split} 
\end{equation}
\begin{equation}
\begin{split}
tr(\bm{R^2})  = \sum_{n_1=1}^{N}\sum_{n_2=1}^{N}|\bm{R}(n_1,n_2)|^2 .
\end{split} 
\end{equation}

\par Then, the EDoF can be expressed as
\begin{equation}
\begin{split}
\beta & = \dfrac{(tr(\bm{R}))^2}{tr(\bm{R}^2)}=\dfrac{|\sum_{n=1}^{N}\bm{R}(n,n)|^2}{\sum_{n_1=1}^{N}\sum_{n_2=1}^{N}|\bm{R}(n_1,n_2)|^2} ,
\end{split} 
\end{equation}
where the analytic expression of $(tr(\bm{R}))^2$ and $tr(\bm{R}^2)$ can be expressed as

\begin{equation}
\begin{split}
(tr(\bm{R}))^2 & = |\sum_{n=1}^{N}\bm{R}(n,n)|^2 =\dfrac{1}{(4\pi)^4} \\
& |\sum_{n=1}^{N}\sum_{m=1}^{M}\dfrac{1}{(x_R)^2 + (y_m-y_n)^2 +(z_m-z_n)^2}|^2 ,
\end{split} 
\end{equation}
where $(x_R)^2 + (y_m-y_n)^2 +(z_m-z_n)^2= (x_R)^2 + (y_R+(k(m)-M_h -i(n)+N_h)\Delta l)^2 + (z_R+(w(m)-M_v -v(n)+N_v)\Delta l)^2$.

In general, the size of the transmitting antenna array is less than the distance $d$ between the receiving and transmitting terminals, so $tr(\bm{R}^2)$ can be written as
\begin{equation}
\begin{split}
tr(\bm{R}^2) &= \sum_{n_1=1}^{N}\sum_{n_2=1}^{N}|\bm{R}(n_1,n_2)|^2 \\
 &=\sum_{n_1=1}^{N}\sum_{n_2=1}^{N}|\sum_{m=1}^{M}\dfrac{exp(-jk_0(d_{m,n_1}-d_{m,n_2}))}{(4\pi)^2 d_{m,n_1}d_{m,n_2}}|^2\\
& = \sum_{n_1=1}^{N}\sum_{n_2=1}^{N}|\sum_{m=1}^{M}\dfrac{exp(-\dfrac{jk_0}{x_R}f(d))}{(4\pi x_R)^2}|^2 ,
\end{split} 
\end{equation}
where $f(d) = y_m(y_{n_1}-y_{n_2})+z_m(z_{n_1}-z_{n_2})$, which can be modified as

\begin{equation}
\begin{split}
f(d) &= y_m(y_{n_1}-y_{n_2})+z_m(z_{n_1}-z_{n_2})\\
&= (y_R+(k(m)-M_h)\Delta l)(i(n_1)-i(n_2))\Delta l\\
&+(z_R+(w(m)-M_v)\Delta l)(v(n_1)-v(n_2))\Delta l.
\end{split} 
\end{equation}

\par Therefore, the closed-form analytical model of the EDoF can be written as	
	
\begin{equation}
\begin{split}
\beta & = (\dfrac{tr(\bm{R})}{||\bm{R}||_F})^2 = \dfrac{(tr(\bm{R}))^2}{tr(\bm{R}^2)}\\
&= \dfrac{|\sum_{n=1}^{N}\sum_{m=1}^{M}\dfrac{1}{q(d)}|^2}{\dfrac{1}{( x_R)^4}\sum_{n_1=1}^{N}\sum_{n_2=1}^{N}|\sum_{m=1}^{M}exp(-\dfrac{jk_0}{x_R}f(d))|^2},
\end{split} 
\end{equation}
where the $q(d)$ and $f(d)$ can be expressed as

\begin{equation}
\begin{split}
q(d)& = (x_R)^2 + (y_R+(k(m)-M_h -i(n)+N_h)\Delta l)^2 \\
&+ (z_R+(w(m)-M_v -v(n)+N_v)\Delta l)^2,\\
f(d)&= (y_R+(k(m)-M_h)\Delta l)(i(n_1)-i(n_2))\Delta l\\
&+(z_R+(w(m)-M_v)\Delta l)(v(n_1)-v(n_2))\Delta l.
\end{split} 
\end{equation}





\subsection{Channel Capacity}

\subsubsection{Expression}
\par The $\bm{R}$ is a unit matrix, and all the subchannels are independent. The channel capacity can be expressed as
\begin{equation}
\begin{split}
C & = r {\rm log}_2{(1+\dfrac{P}{rN_0})},
\end{split} 
\end{equation}
where $\dfrac{P}{N_0}$ is the transmit signal-to-noise ratio (SNR), $r$ is the rank of the correlation matrix, which is the degree of freedom of the correlation matrix $\bm{R}$.

\par The channel capacity of near-field MIMO is expressed by introducing (8), which can be written as

\begin{equation}
\begin{split}
C & = \beta {\rm log}_2{(1+\dfrac{P}{rN_0})},
\end{split} 
\end{equation}
where $\beta  \in [1, r]$ is the EDoF, i.e., the equivalent number of SISO systems.

\subsubsection{Closed-Form Model}

\par Therefore, the closed-form analytical model of channel capacity can be written as

\begin{equation}
\begin{split}
C & = \dfrac{|\sum_{n=1}^{N}\sum_{m=1}^{M}\dfrac{1}{q(d)}|^2}{\dfrac{1}{( x_R)^4}\sum_{n_1=1}^{N}\sum_{n_2=1}^{N}|\sum_{m=1}^{M}exp(-\dfrac{jk_0}{x_R}f(d))|^2} \\&\cdot {\rm log}_2{(1+\dfrac{P}{rN_0})}\\
& = (x_R)^4\dfrac{|\sum_{n=1}^{N}\sum_{m=1}^{M}\dfrac{1}{q(d)}|^2}{\sum_{n_1=1}^{N}\sum_{n_2=1}^{N}|\sum_{m=1}^{M}exp(-\dfrac{jk_0}{x_R}f(d))|^2} \\&\cdot {\rm log}_2{(1+\dfrac{P}{rN_0})},
\end{split} 
\end{equation}
where $q(d)=(x_R)^2 + (y_R+(k(m)-M_h -i(n)+N_h)\Delta l)^2 + (z_R+(w(m)-M_v -v(n)+N_v)\Delta l)^2$, and $f(d)= (y_R+(k(m)-M_h)\Delta l)(i(n_1)-i(n_2))\Delta l+(z_R+(w(m)-M_v)\Delta l)(v(n_1)-v(n_2))\Delta l$.




\section{Numerical Analysis of Near-Field Performance at Mid-Band}

\subsection{Verification of Measurement Experiment}

\par To verify the closed-form model of near-field channel capacity, we conducted an indoor channel measurement in the mid-band. The measurement platform is extended on the experimental system in [21-22]. As shown in Fig. 2, the near-field measurement is carried out at 13 GHz in the indoor-office environment. In Fig. 2 and Table \uppercase\expandafter{\romannumeral1}, the mid-band and mmWave massive MIMO channel measurement platform based on high-speed electronic switching switch can complete massive MIMO channel measurement within one hundred milliseconds, making it a reality to measure the comprehensive characteristics of mid-band MIMO channel based on TDM-MIMO mode. Combined with pseudo-random (PN) sequence signal and time division multiplexing antenna switching technology, the channel data is collected. The PN sequence length is 511. The SNR is set to 70 dB, and $y_R=z_R=0$. The UPA antenna at the transmitting end is a 128-element MIMO array based on the high-precision measurement platform, which is positioned in the front of the room. The Rx with eight array elements is located in the aisle. The Rx ends are distributed in the near-field range, and the nearest distance is 1 m between the receiving and transmitter end.

\begin{figure}[!htbp]
	\xdef\xfigwd{\columnwidth}
	\setlength{\abovecaptionskip}{0.1 cm}
	\centering
	\begin{tabular}{cc}
		\includegraphics[width=4.4cm,height=3.0cm]{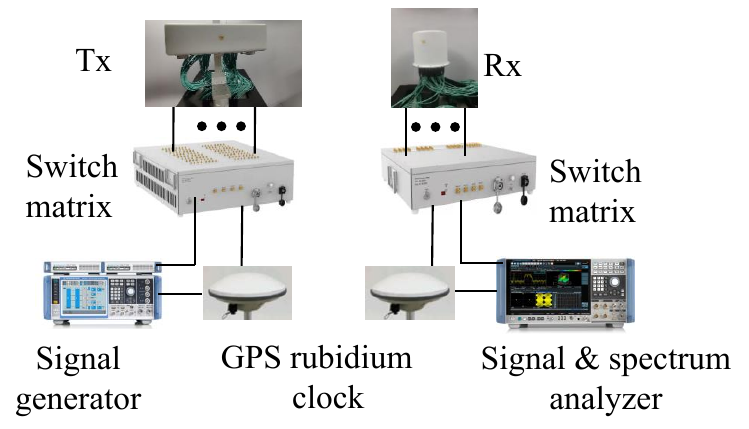} \hspace{-5mm}
		& \includegraphics[width=4.4cm,height=3.4cm]{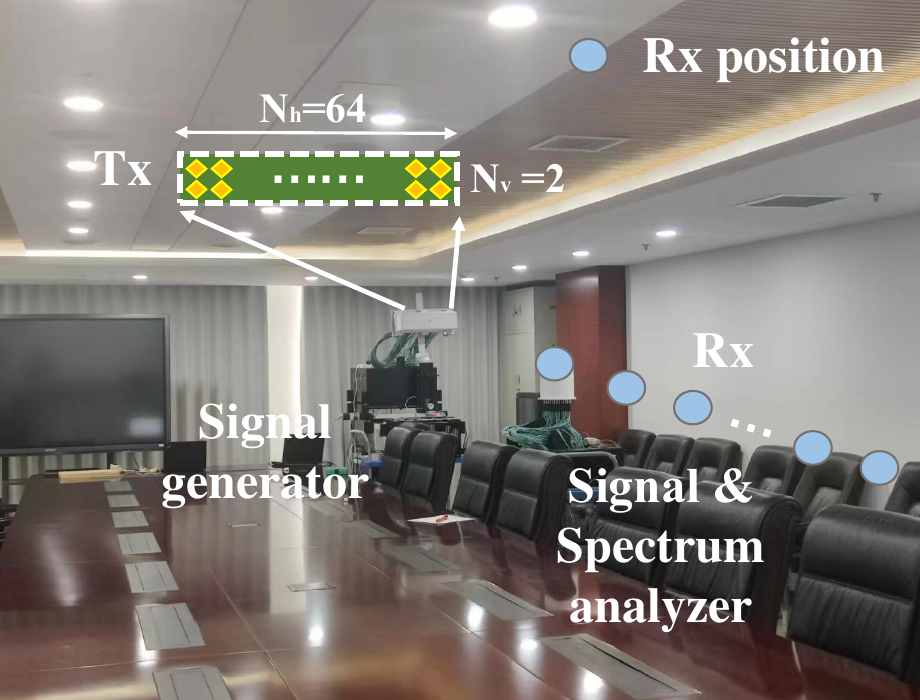}\\
		{\footnotesize\sf (a)} &	{\footnotesize\sf (b)} \\		
	\end{tabular}
	\caption{Mid-band near-field MIMO channel measurement experiment. (a) Schematic diagram. (b) Indoor-office environment.}
	\label{fig:Figure26}
\end{figure}

\begin{table}[!htbp]
	\centering
	\caption{Channel Measurement Parameter Setting}
	\setlength{\tabcolsep}{0.1 mm}
	\label{my-label}
	\renewcommand{\arraystretch}{1.8}
	\setlength{\tabcolsep}{11 mm}
	\begin{tabular}{c|c}
		\hline \hline
		\textbf{Parameter}    & \textbf{Value}  \\ \hline
		Frequency [GHz]    & 13  \\ \hline
		$N_h\times N_v$    & $64\times 2$  \\  \hline
		$M_h\times M_v$     & $4\times 2$\\  \hline
		Antenna type  & UPA  \\  \hline
		Antenna element spacing $\Delta l$   & $\lambda_0/2$  \\  \hline
		SNR [dB]  & 70  \\  \hline  
		\hline
	\end{tabular}
\end{table}

\begin{figure}[!htbp]
	\xdef\xfigwd{\columnwidth}
	\setlength{\abovecaptionskip}{0.1 cm}
	\centering
	\begin{tabular}{cc}
		\includegraphics[width=4.4cm,height=3.6cm]{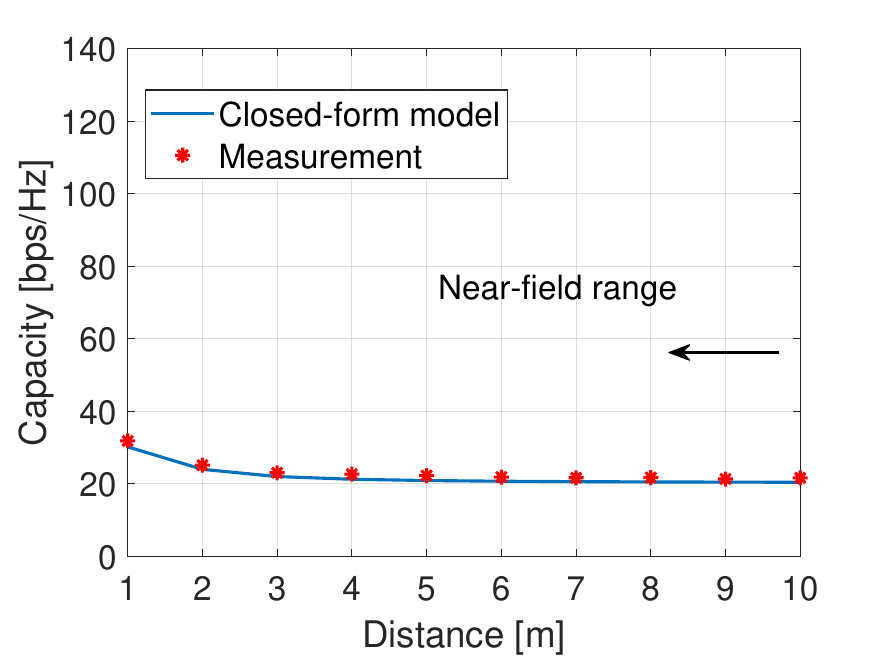} \hspace{-5mm}
		& \includegraphics[width=4.4cm,height=3.6cm]{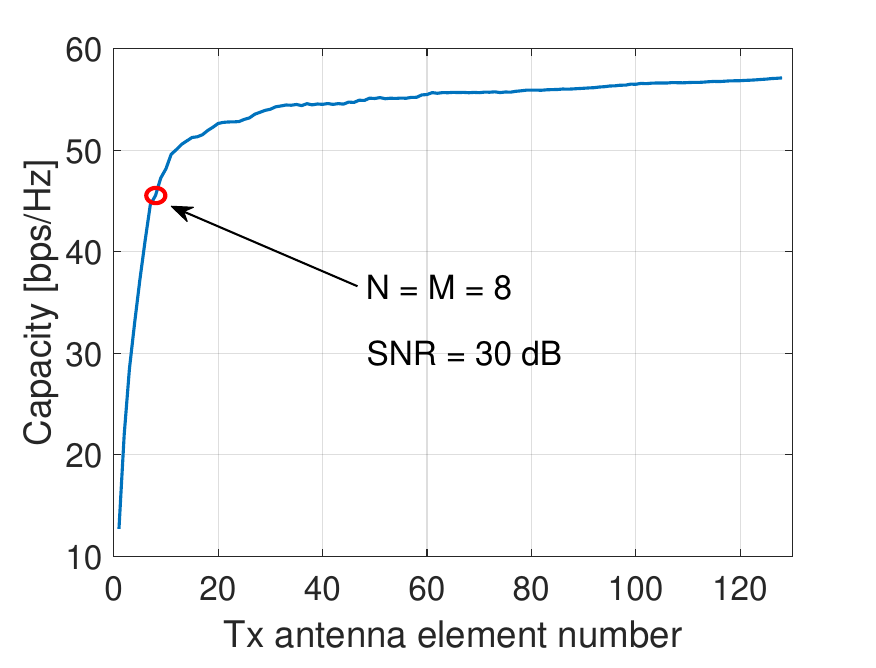}\\
		{\footnotesize\sf (a)} &	{\footnotesize\sf (b)} \\		
	\end{tabular}
	\caption{Measurement results and verification of channel capacity. (a) Versus the communication distance. (b) Versus the number of Tx antenna element.}
	\label{fig:Figure6}
\end{figure}

\par In Fig. 3(a), as the distance between receiving and transmitter array centers increases, the channel capacity of UPA-type MIMO systems can be observed to decrease continuously, and the rate of channel capacity reduction becomes slower. The theoretical channel capacity based on EDoF is basically consistent with the measurement results. The closed-form expression of channel capacity can basically be used to describe the near-field channel capacity within a certain distance, and more importantly, the model is easier to interpret and process. In Fig. 3(b), compared with the far-field characteristics, the channel capacity of near-field MIMO system is still related to the minimum number of antennas at the receiving and transmitter end. The results show that the near-field MIMO capacity gain is limited in the actual communication configuration.

\subsection{Numerical Analysis of Channel Capacity }

In this section, the factors affecting near-field capacity gain are discussed. We can set $f = 13$ GHz, $SNR = 70$ dB, $y_R=z_R= 0$.

\begin{figure}[!htbp]
	\xdef\xfigwd{\columnwidth}
	\setlength{\abovecaptionskip}{0.1 cm}
	\centering
	\begin{tabular}{cc}
		\includegraphics[width=4.4cm,height=3.6cm]{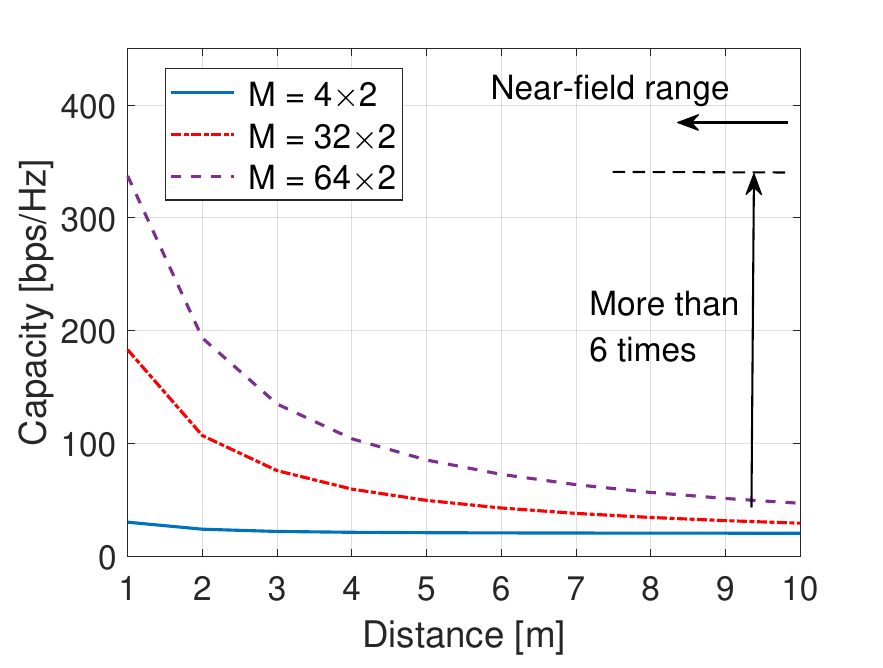} \hspace{-5mm}
		& \includegraphics[width=4.4cm,height=3.6cm]{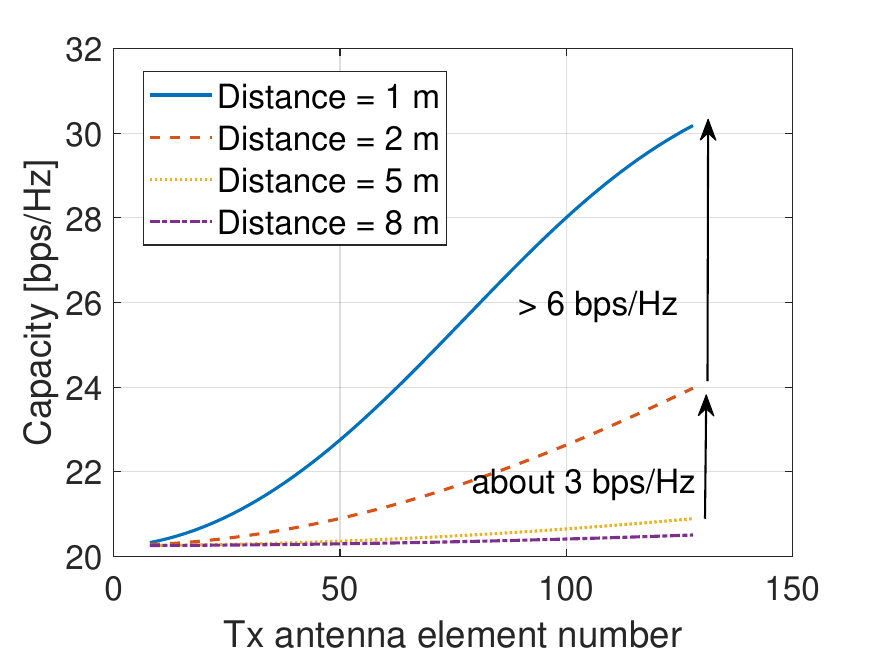}\\
		{\footnotesize\sf (a)} &	{\footnotesize\sf (b)} \\		
	\end{tabular}
	\caption{ Numerical results of channel capacity. (a) Versus the communication distance with different Rx antenna element number $M={8, 64, 128}$, and Tx antenna element number $N=64\times 2$. (b) Versus the communication distance and Tx antenna element number, and Rx antenna element number $M=4\times 2$.}
	\label{fig:Figure6}
\end{figure}

\par Interestingly, as shown in Fig. 4(a), it can be observed that with the size of the receiver array increasing, the channel capacity gain becomes more and more obvious in the near position. Therefore, the near-field capacity gain is relatively obvious when the large-scale MIMO is adopted at both receiving and transmitter end, but considering that the antenna array of user is relatively small in the actual communication system, the near-field capacity gain may not be obtained in this configuration. In Fig. 4(b), with the increase of the number of antennas, the near-field channel capacity gain increases when the position of receiving and transmitter end is fixed. As the receiver and transmitter get closer, the channel capacity gain becomes more obvious.


\section{Conclusion}

\par In this letter, the capacity has been investigated and verified based on the mid-band near-field channel measurement. The EDoF and channel capacity are analyzed theoretically, and their closed-form analytical expressions are derived. The near-field channel capacity gain is relatively obvious when large-scale MIMO is adopted at both receiving and transmitter ends. As the distance increases, the channel capacity of UPA-type MIMO systems can be observed to decrease continuously. The near-field channel capacity gain is limited in the actual communication system with the small antenna array of user. Based on the mid-band near-field experimental study, the upper limit of the near-field channel capacity is also related to the minimum number of antennas at the receiving and transmitter end.

\ifCLASSOPTIONcaptionsoff
\newpage
\fi

\end{document}